\documentclass{jpsj3}

\makeatletter
\newcommand\newblock{\hskip .11em\@plus.33em\@minus.07em}
\makeatother
\usepackage{float}
\usepackage{booktabs}
\usepackage{txfonts}
\usepackage{authblk}
\usepackage{url}
\usepackage{xcolor}

\usepackage[utf8]{inputenc}
\usepackage[T1]{fontenc}
\usepackage{qcircuit}
\usepackage{braket}

\usepackage{graphicx}

\usepackage{bm}

\title{Quadratic Unconstrained Binary Formulation for Traffic Signal Optimization on Real-World Maps}
\author{Reo Shikanai$^{1,2*}$, Masayuki Ohzeki$^{1,2,3}$, and Kazuyuki Tanaka$^{}$}

\affil[1]{Graduate School of Information Sciences, Tohoku University, Sendai 980-8579, Japan}
\affil[2]{Sigma-i Co., Ltd., Tokyo 108-0075, Japan}
\affil[3]{Department of Physics, Tokyo Institute of Technology, Tokyo 152-8551, Japan}

\abst{
The D-Wave quantum annealing machine can quickly find the optimal solution for quadratic unconstrained binary optimization (QUBO).
One of the applications where the use of quantum annealing is desired is in problems requiring rapid calculations.
One such application is the traffic signal optimization.
Several studies have used quantum annealing; however, they are formulated in relatively unrealistic settings, such as only crossroads on a map.
We propose a different formulation of QUBO that can also deal with T-junctions and multi-forked roads.
The simulation of urban mobility (SUMO) was used to validate the efficiency of our approach and verify the feasibility of real-time control using geographical information data that were very similar to the real world.
Our model could reduce the waiting time at red lights for vehicles.
In addition, we compared our results with those of the Gurobi Optimizer to confirm whether the D-Wave machine could find the ground state.
Unfortunately, our results show that the D-Wave machine could not find the optimal solution and was slower than the Gurobi Optimizer in computation time.
}

\inst{}

\begin{document}
\maketitle

\section{Introduction}
The quantum annealing machine, developed by D-Wave Systems Inc., has attracted significant attention as the first commercially available device in quantum computing. 
It solves combinatorial optimization problems using the quantum annealing (QA) protocol.
QA is a heuristic algorithm for solving combinatorial optimization problems using quantum fluctuations \cite{kadowaki_quantum_1998}.
Quantum annealers have been validated in various applications, exhibiting their potential effect in several fields, such as traffic flow optimization \cite{neukart2017traffic,hussain2020optimal,inoue2021traffic, Ohzeki2024}, nurse-scheduling problems \cite{ikeda_application_2019},
finance \cite{rosenberg2016solving,  venturelli2019reverse}, logistics \cite{mugel_dynamic_2022}, manufacturing \cite{venturelli2016quantum, Yonaga2022, Haba2022}, material pre-processing \cite{Tanaka2023}, searching for nontrivial candidate \cite{Doi2023}, marketing \cite{nishimura2019item}, maze generation \cite{Ishikawa2023}, steel manufacturing \cite{Yonaga2022}, and decoding problems \cite{IdeMaximumLikelihoodChannel2020, Arai2021code}.
Fast solvers have also been proposed by integrating with classical methods of operations research as in the literature \cite{Hirama2023, takabayashi2024}.
Model-based Bayesian optimization using quantum annealing has also been proposed \cite{Koshikawa2021, morita2023}.
A comparative study of QA was performed with benchmark tests to solve optimization problems \cite{Oshiyama2022}. 
Because environmental effects cannot be avoided, the D-Wave machine is sometimes considered a simulator of quantum many-body dynamics \cite{Bando2020, Bando2021, King2022}. 
In addition, QA has been applied as an optimization algorithm in machine learning \cite{neven2012qboost,khoshaman2018quantum,o2018nonnegative, Amin2018, Kumar2018, Arai2021, Sato2021, Urushibata2022, hasegawa2023}.

In this study, QA is used for traffic signal optimization.
One role of traffic signals is to improve the traffic flow\cite{papageorgiou_review_2003}, which can help vehicles reach their destinations faster as well as reduce traffic pollution\cite{zhang_air_2013} such as exhaust emissions and noise. 
Traffic-sensitive control is currently used at intersections with particularly heavy traffic in Japan\cite{miho_2004}. 
This method involves transmitting the real-time traffic volume to a control center, which calculates and sends appropriate signal lighting times back. 
Thus, traffic congestion can be eliminated more flexibly than fixed-cycle control, which constantly repeats the same lighting time. 
However, the disadvantage is that the number of calculations increases exponentially as the number of intersections to be controlled increases.
High-speed computation is necessary as real-time calculations are required for traffic signal optimization.
We focus on the potential of the QA machine manufactured by D-Wave Systems in this context.
Although several previous studies have used D-Wave machine, they have not been formulated and tested in realistic settings.
The study by Inoue \textit{et al.}\cite{inoue_traffic_2021} defined only two intersection states: vehicles can proceed in the north-south or east-west direction.
Thus, this method cannot deal with right-turn-only lanes or multi-forked roads.
The study by Hussain \textit{et al.}\cite{hussain_optimal_2020} was more realistic because it defined six intersection states, including right-turn signals.
However, all intersections were assumed to be cross-streets, which is impractical.
In addition, a simulation was performed on a square grid, and the road length and other settings deviated from those in the real world.
The quadratic terms in the cost function were intended to make it easier for vehicles to pass through intersections continuously, but the simulation results were almost the same as the cost function without the quadratic terms.
In that study, the coefficient of the quadratic term depended on the expected number of vehicles passing through the intersection.
We propose the cost function that removes such uncertainties from the quadratic term and incorporates a quadratic term that depends only on road information such as road length, speed limit, and pattern compatibility between two intersections.
This means that these coefficients are calculated once and do not need to be updated, which reduces the time required to formulate the cost function.

The number of qubits available on D-Wave machines has increased with their development.
For example, D-Wave 2000Q, introduced in 2017, offers approximately 2,000 qubits.
The Advantage system, introduced in 2020, has more than 5,000 qubits, providing a significant leap in computational power compared with the previous model.
The version of these machines and the number of qubits directly affect their computational speed \cite{willsch_benchmarking_2022}.
In addition, the physical qubits of the D-Wave machine are not all directly connected.
Thus, if the variables in a problem do not match the connection structure of the physical qubits, they must be embedded to represent the problem on the machine.
This process is referred to as minor embedding, and finding such embeddings in the hardware graph of D-Wave machines is generally an NP-hard problem \cite{robertson_graph_1995}. 
Embeddings are pre-designed to align with specific graph structures such as cliques, simplifying the embedding process significantly.
However, identifying minor embeddings in damaged or irregular graphs remains an NP-complete problem \cite{lobe_minor_2021}.
This embedding process may require additional qubits and connections, thereby increasing the size and complexity of the problem.
Specifically, a single logical variable is represented by multiple qubits, and this group is known as a chain.
The qubits in the chain must have the same value, which requires strong ferromagnetic interactions between the qubits.
Discrepancies within the chain, known as chain breaks, are typically resolved through post-processing methods such as majority voting to ensure uniformity. 
Determining the appropriate strength of ferromagnetic interactions between qubits is challenging; if not set correctly, chain breaks can negatively impact the solution search\cite{venturelli_quantum_2015}.
Methods for mitigating this effect and post-processing methods have also been studied \cite{vinci_quantum_2015}.
This study used the default setting for post-processing for D-Wave machines, namely majority vote.
In addition to these hardware characteristics, the D-Wave machine has the property that it can only solve quadratic unconstrained binary optimization (QUBO). Therefore, we need to formulate the cost function in QUBO form.

The D-Wave machine does not always output the optimal solution owing to hardware restrictions such as the chain and environmental noise. \cite{kadowaki_experimental_2019}
Therefore, the results of the QA are compared with those of the Gurobi Optimizer \cite{gurobi} in this study.
This commercial solver outputs an optimal solution, and we examine the proximity of this solution to that obtained from QA.

The remainder of this paper is organized as follows: 
In the following section, we describe the formulation used to eliminate congestion. We introduce the specific experimental setup and compare the results from various perspectives. 
Finally, we summarise our study and the future direction of our approach.

\section{Methods}
The problem set for intersections is described, and the differences from existing methods\cite{hussain_optimal_2020}\cite{inoue_traffic_2021} are outlined in this section. 
Subsequently, the QUBO for reducing traffic congestion is formulated.

\subsection{Preliminaries}
\begin{figure}[htbp]
\centering
\includegraphics[width=\columnwidth, keepaspectratio]{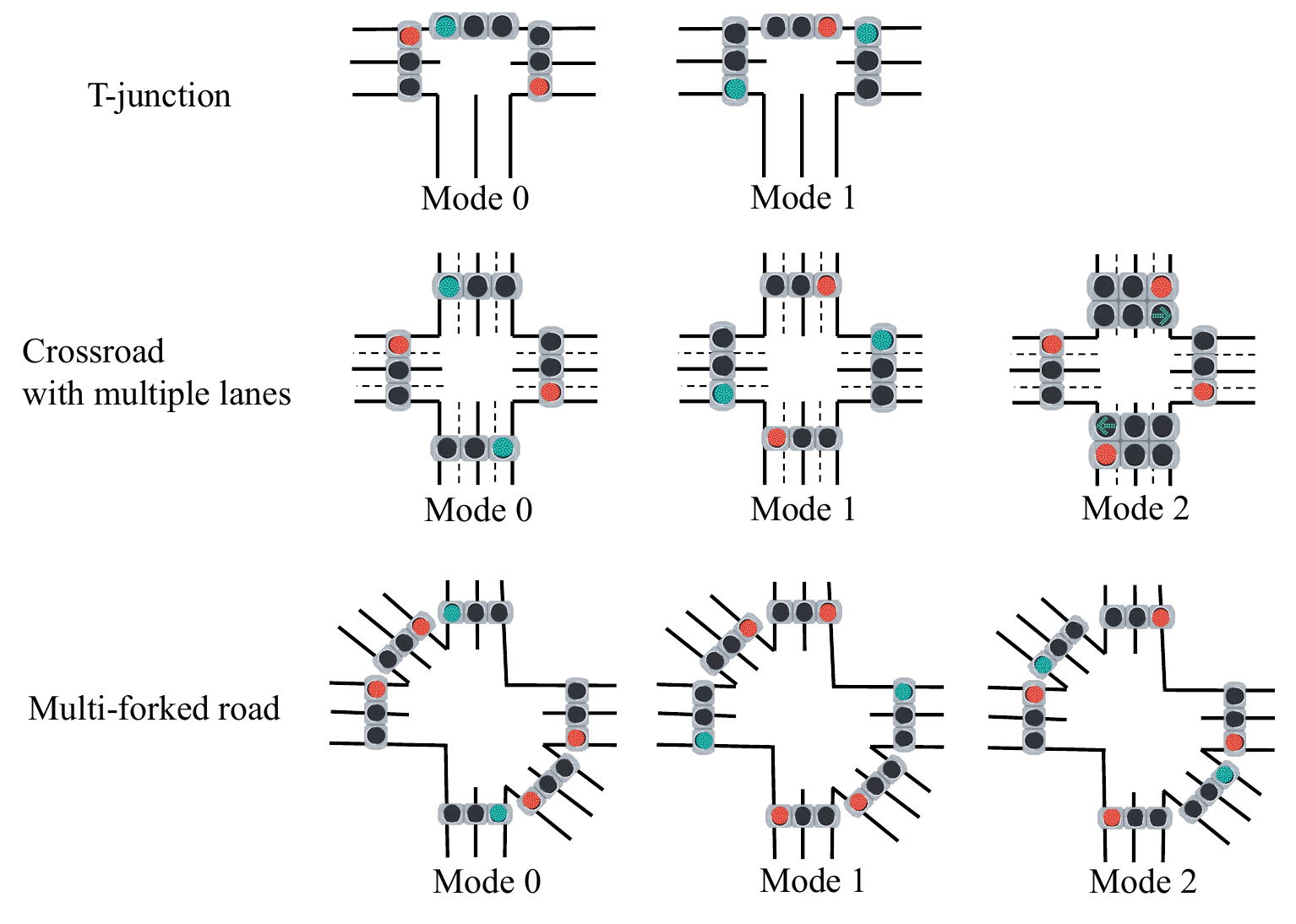}
\caption{
(Color online)
Example definition of 'mode'.
Various numbers and types of modes at different intersections.
For example, at a crossroads with multiple lanes, cars in the right-turn-only lane can proceed when in mode 2.}
\label{fig:sample_modes}
\end{figure}

We introduce a variable with several states at each intersection.
For example, as shown in Figure \ref{fig:sample_modes}, a T-junction has two states, each referred to as a `mode'.
A right-turn lane exists at a crossroad with multiple lanes.
A multi-forked road has a mode where traffic in the diagonal direction can proceed.
Thus, the types and the number of possible modes vary among intersections.
These settings accurately reflect real intersections.
We define $x_{im}=1$ when mode $m$ is selected at the $i$-th intersection and $0$ otherwise.

\subsection{Formulation}
We design the cost function to reduce the time each vehicle spends stopped at a red light, which we refer to as the \textit{waiting time}. The \textit{total waiting time} is then defined as the sum of all individual \textit{waiting times} for each vehicle. However, to design a cost function that directly minimizes this \textit{total waiting time}, it would be necessary to obtain the \textit{waiting time} for each vehicle, which is challenging in practice. This is because it requires constant monitoring of all vehicles to measure their individual \textit{waiting times}. 
Therefore, we create a cost function to reduce the \textit{total waiting time} indirectly. Specifically, the cost function is decomposed into three components, which are linearly combined to form the final cost function to be solved. In this section, each component is described in detail.

The first element maximizes the number of vehicles passing through each intersection.
\begin{equation}
H_1(\vec{x}):=- \sum_{i=1}^{N} \sum_{m \in M_i} C_{i m} x_{i m},
\end{equation}
where $C_{im}$ is the number of vehicles that can pass through the $i$-th intersection if mode $m$ is also selected, $N$ is the number of intersections, and $M_{i}$ is the set of modes of the $i$-th intersection.
$C_{im}$ is scaled by division by $\max_{im} C_{im}$ such that the coefficient of $H_{1}$ does not become excessively large as the traffic volume increases.
We define $C$ as follows: $C=\{C_{im}|\forall i \in \{ 0,...,N-1 \}, \forall m \in \{ 0,...,N_{i}-1 \} \}$.

Mode selection by minimizing $H_1$ allows most vehicles to pass at each intersection.
However, it does not consider the mode of adjacent intersections at all.
Traffic congestion cannot be resolved if the next intersection is not continuously passed. 
Thus, we introduce Equation \ref{equ:second_cost} to allow vehicles to pass continuously through the intersections.
\begin{equation}
\label{equ:second_cost}
H_2(\vec{x}):=-\beta \sum_{i=1}^{N} \sum_{m \in M_i} \sum_{j \in N_i} \sum_{n \in M_j} B_{i j} R_{i m, j n} x_{i m} x_{j n},
\end{equation}
where $\beta$ is a real-valued hyperparameter, $N_i$ is the set of adjacent intersections of the $i$-th intersection, $B_{ij}$ is the degree of connectivity between $i$ and $j$, and is a positive real number. 
The parameter \( R_{im, jn} \) represents the "compatibility" when the \( i \)-th intersection takes the \( m \)-th mode, and the adjacent \( j \)-th intersection takes the \( n \)-th mode. We can encourage the selection of such patterns by setting higher compatibility for situations where a vehicle can consecutively pass through two adjacent intersections. Specifically, \( R_{im, jn} \) is defined as follows:

\begin{equation}
\label{equ}
R_{im, jn} := \begin{cases}
2 & \text{If it is possible to move from } i \text{ to } j \text{ and from } j \text{ to } i, \\
1 & \text{If it is possible to move from } i \text{ to } j \text{ or from } j \text{ to } i., \\
0 & \text{If it is not possible to move from either } i \text{ to } j \text{ or } j \text{ to } i..
\end{cases}
\end{equation}

\begin{figure}[htbp]
\centering
\includegraphics[width=0.7\columnwidth, keepaspectratio]{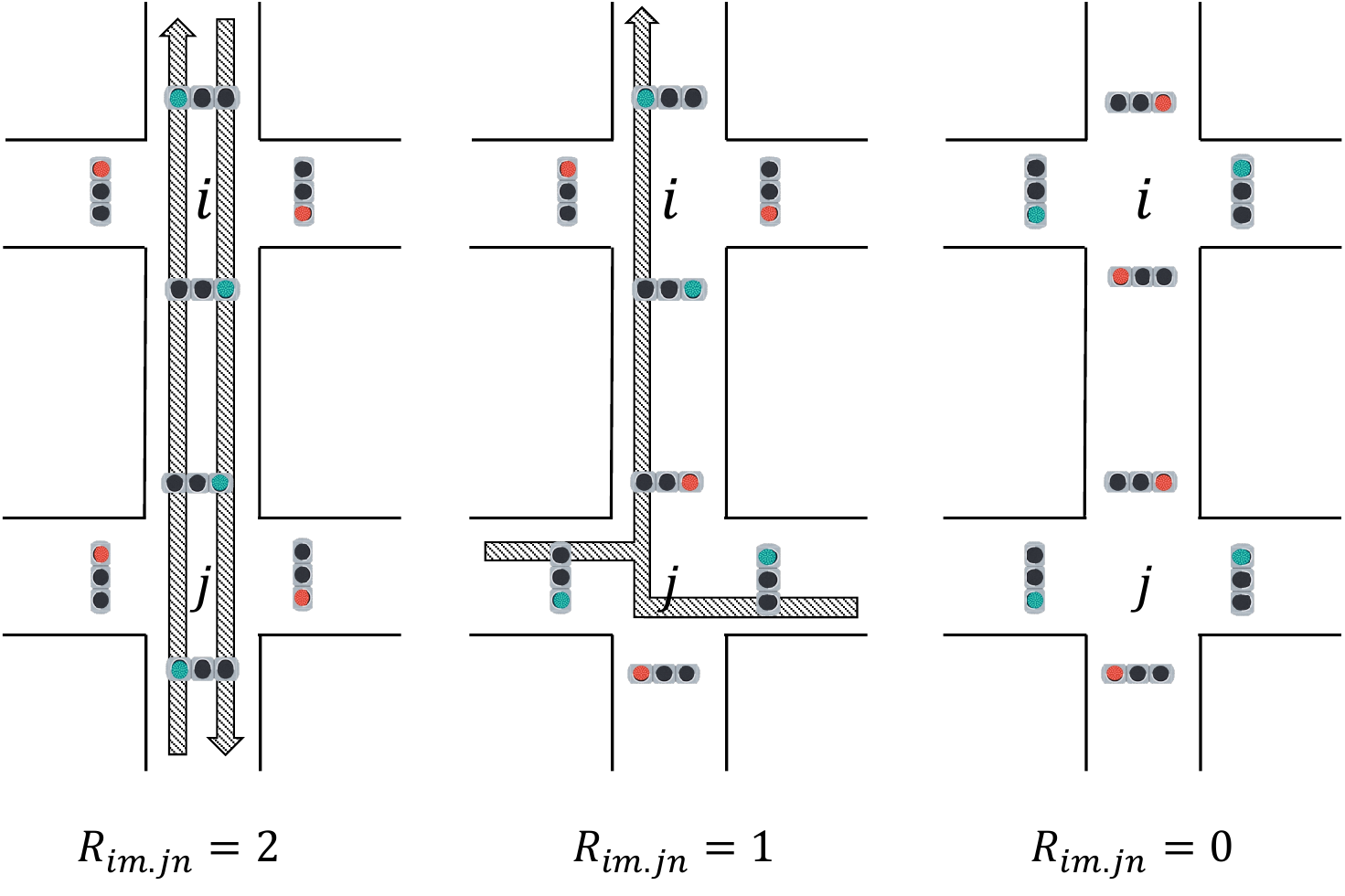}
\caption{
(Color online)
Examples of \( R_{im, jn} \). In the left example, when the north-south direction has a green light at both the \( i \)-th and \( j \)-th intersections, a vehicle can consecutively pass through both intersections in both directions, from \( i \) to \( j \) and from \( j \) to \( i \). Thus, \( R_{im, jn} = 2 \). 
In the middle example, if the east-west direction has a green light at the \( j \)-th intersection, a vehicle can pass consecutively only in the direction from \( j \) to \( i \). Hence, \( R_{im, jn} = 1 \).
Finally, in the right example, when a vehicle cannot pass consecutively in either direction from \( i \) to \( j \) or from \( j \) to \( i \), \( R_{im, jn} = 0 \).
}
\label{fig:sample_relationship}
\end{figure}

For example, as shown on the left side of Figure \ref{fig:sample_relationship}, when the north-south direction has a green light at the \( i \)-th intersection and the north-south direction also has a green light at the \( j \)-th intersection, vehicles can consecutively pass through both intersections from \( i \) to \( j \) and from \( j \) to \( i \). Thus, \( R_{im, jn} = 2 \).
In contrast, as illustrated in the middle of Figure \ref{fig:sample_relationship}, if the east-west direction has a green light at the \( j \)-th intersection, vehicles can pass consecutively only in the direction from \( j \) to \( i \). Hence, \( R_{im, jn} = 1 \).
Finally, as shown on the right side of Figure \ref{fig:sample_relationship}, when vehicles cannot pass consecutively in either direction from \( i \) to \( j \) or from \( j \) to \( i \), \( R_{im, jn} = 0 \).
In this way, \( R_{im, jn} \) can be defined for any pair of adjacent intersections, even if they form a T-intersection or a multi-forked intersection.

It is expected that $R_{im, jn}$ will make it easier to select a pattern of modes allowing vehicles to pass through the intersection continuously.
On the other hand, if it takes longer to arrive at intersection $j$ from intersection $i$, the contribution of $R_{im, jn}$ should be weakened.
Therefore, we define $B_{ij}$ as follows.
\begin{equation}
\label{equ:B}
B_{i j} := \frac{1}{\text{Time to arrive from } i \text{ to } j } = \frac{{\text{Speed limit between } i \text{ and } j }} {\text{Distance between } i \text{ and } j },
\end{equation}
where $B_{ij}$ is scaled by dividing it by $\max_{ij} B_{ij}$, and the unit is dimensionless.
Multiplying $B_{ij}$ by $R_{im, jn}$ adjusts the compatibility between intersections to a strength that depends on travel time.

Finally, penalty terms are introduced to satisfy the one-hot constraint in Equation \ref{equ:third_cost}.
\begin{equation}
\label{equ:third_cost}
H_3(\vec{x}):=\gamma \sum_{i=1}^{N}\left(\sum_{m \in M_i} x_{i m}-1\right)^2,
\end{equation}
where $\gamma$ denotes a real-valued hyperparameter. 
The final cost function to be solved is as follows:
\begin{equation}
\label{equ:final_cost}
\min_{\vec{x} \in \{0,1\}^{M}} \quad H_1(\vec{x})+H_2(\vec{x})+H_3(\vec{x}),
\end{equation}
where $M=\sum_{i=1}^{N} M_{i}$.
Traffic congestion-relieving mode are expected to be selected by finding $\vec{x}$ that minimizes the cost function.
In the following section, we validate whether the optimal solution to Equation \ref{equ:final_cost} can alleviate traffic congestion within the simulator.

\subsection{Problem settings}
\begin{figure}[htbp]
\centering
\includegraphics[width=\columnwidth, keepaspectratio]{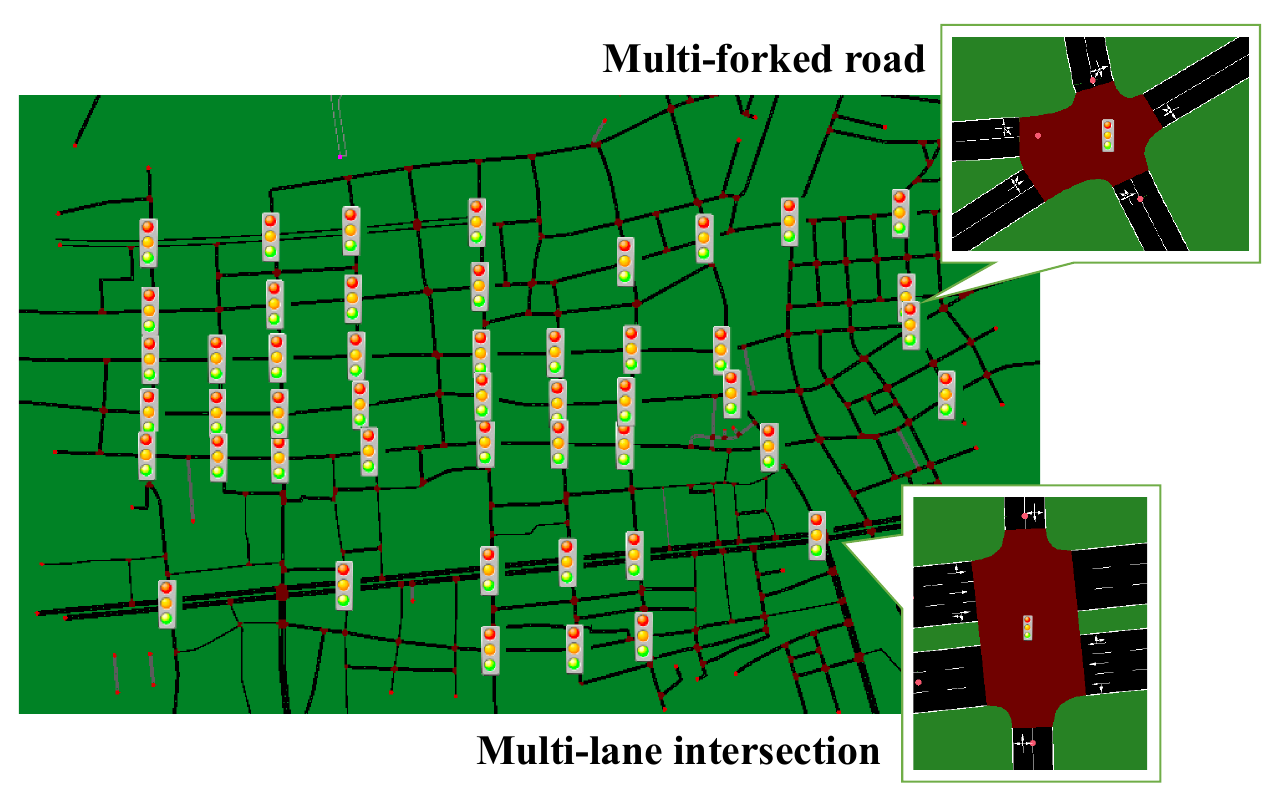}
\caption{
(Color online)
We used a map for simulation, utilizing a portion of Aomori City in Japan. This map includes multi-forked roads and multi-lane intersections. Intersections with traffic signals are marked with a traffic light symbol.
}
\label{fig:map}
\end{figure}

SUMO\cite{SUMO2018} is an open software program that allows users to download terrain data from open street maps and move vehicles on them. The map used is shown in Figure \ref{fig:map}.
For this experiment, we used a section of Aomori City, located in the northernmost region of Honshu, the main island of Japan. Aomori City is a relatively high-traffic area in Aomori Prefecture, and the map includes multi-forked roads and multi-lane intersections. Signalized intersections are marked with a traffic signal symbol, and 48 such intersections are mapped. Since each intersection has two or three operational modes, the number of binary variables is 100.

\begin{figure}[htbp]
\centering
\includegraphics[width=\columnwidth, keepaspectratio]{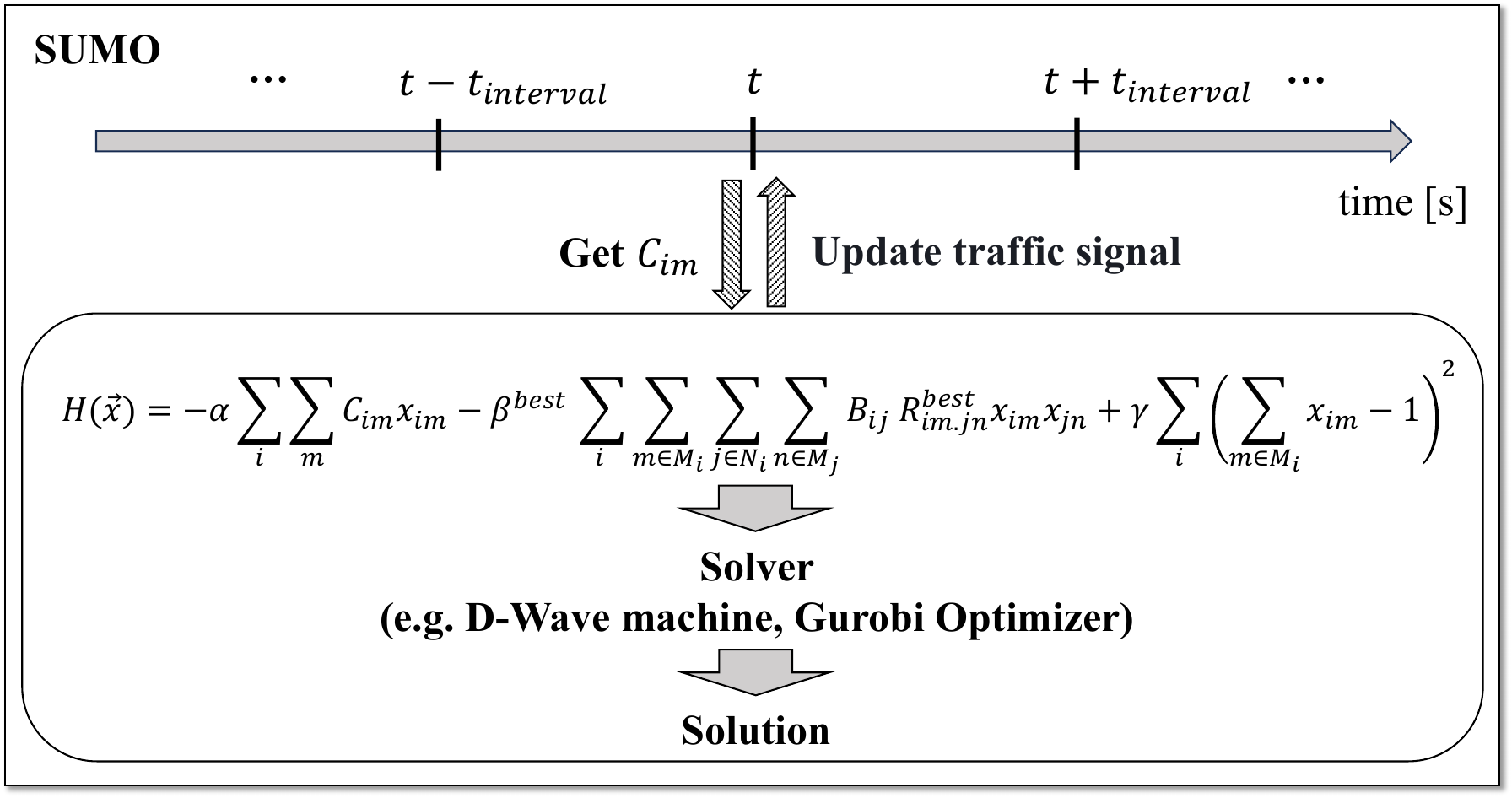}
\caption{
(Color online)
Overview of the simulation.
Optimization is performed every $t_{interval}$.
At time $t$, the traffic volume information, $C$, is obtained and entered into a cost function.
Next, a solver solves this cost function and updates the state of the traffic lights according to the solution.}
\label{fig:t_sec}
\end{figure}

{
Furthermore, traffic flow data were prepared for five initial vehicle count patterns-200, 300, 400, 500, and 600. The destinations and objectives of each vehicle were randomly assigned, and for each flow level, we created 10 data instances with varying vehicle routes. Figure \ref{fig:t_sec} illustrates the simulation flow for a particular instance. In addition to the initial vehicles, two new cars entered the map every second, while vehicles reaching their destinations were removed from the map.
}

The cost function is minimized at each time interval, \( t_{interval} \). At the time \( t \) designated for optimization, the traffic volume data \( C \) is gathered and applied to the cost function, after which the solver computes the optimal state and updates the signal states accordingly.

\subsection{Solver settings}
 In QA, as implemented in the D-Wave Advantage 6.4 system, critical operational parameters such as $ \texttt{annealing\_time} $, $ \texttt{num\_reads} $, and $ \texttt{anneal\_schedule} $ are pivotal for steering the quantum system towards optimal solutions. The $ \texttt{annealing\_time} $ parameter allows the system to explore the energy landscape effectively. The $ \texttt{num\_reads} $ parameter specifies the number of times quantum annealing is performed. The $ \texttt{anneal\_schedule} $ parameter controls the strength of the quantum fluctuations, and the default setting was used.

We used the $ \texttt{EmbeddingComposite} $ of the D-Wave Ocean SDK, which applies a minor-miner technique \cite{cai_practical_2014} to perform embedding automatically. The specific settings of the D-Wave machine and simulated annealing are listed in Table \ref{table:qa-sa-setting}.

In SA, the scheduling parameter controls the system's transitions from high to progressively lower temperatures. The minimum and maximum energy differences (delta energies) are automatically computed from the Ising graph's interactions in the OpenJij $ \texttt{SASampler} $ default configuration. These values determine the minimum ($ \beta_{\text{min}} $) and maximum ($ \beta_{\text{max}} $) inverse temperatures, which are set by dividing $ \log 2 $ by the maximum delta energy and $ \log 100 $ by the minimum delta energy. The inverse temperature typically decreases geometrically during annealing. The parameter $ \texttt{num\_sweeps} $ represents the total number of Metropolis-Hastings updates performed during the annealing process.

\begin{table}[hbtp]
  \caption{\label{table:qa-sa-setting}Quantum annealing and simulated annealing settings}
  \centering
  \begin{tabular}{lcc}
    \hline
    Property            & Quantum annealing & Simulated annealing \\
    \hline \hline
    Sampler             & Advantage System 6.4        & OpenJij \texttt{SASampler}                       \\
    \texttt{annealing\_time}      & 20 $\mu$s                   & -                               \\
    \texttt{num\_reads}           & 1,000                       & 1,000                           \\
    \texttt{anneal\_schedule}     & Default                     & Default                               \\
    \texttt{num\_sweeps}          & -                           & 1,000                           \\
    \hline
  \end{tabular}
\end{table}

\section{Results}
\begin{figure}[htbp]
\centering
\includegraphics[width=\columnwidth, keepaspectratio]{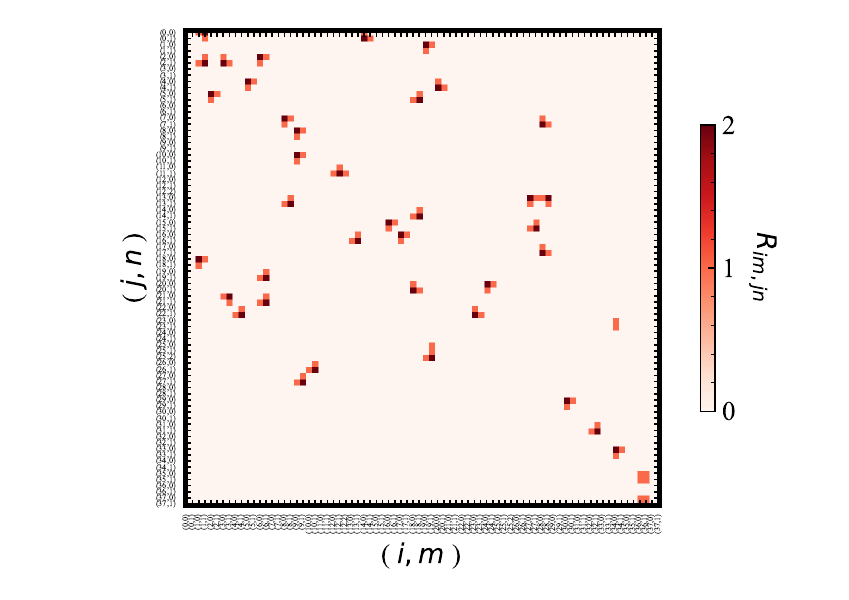}
\caption{
(Color online)
The parameter \( R_{im,jn} \) used in the map. Independent intersections without adjacent intersections are excluded.
}
\label{fig:optimized_R}
\end{figure}

This section presents the method for exploring the parameter $ \beta $ and the results obtained. We also provide a visualization of $ R_{im, jn} $ and discuss the results of simulations conducted using these parameters. As previously mentioned, $ \gamma $ must be set to a high value to satisfy the constraints. Therefore, $ \gamma $ was set to 10, and $ \beta $ was tuned through a grid search within the range of 0 to 0.1 in increments of 0.01. Figure \ref{fig:optimized_R} shows a visualization of $ R_{im, jn} $. Independent intersections without adjacent intersections are excluded. Since these parameters are map-dependent, the same parameters are used in dynamic calculations. This characteristic is advantageous in situations requiring dynamic and fast computation, such as traffic signal optimization, as it eliminates the time needed to calculate the quadratic coefficients of the cost function. The grid search was performed for each traffic flow data set, and the $ \beta $ that minimized the average total \textit{waiting time} across 10 instances was selected. As a result, $ \beta $ values were determined as shown in Table \ref{table:beta_values}, demonstrating that the optimal $ \beta $ varied depending on the initial number of vehicles. The following section presents the results of experiments conducted using these parameters.

\begin{table}[h]
\caption{Values of \(\beta\) for different initial vehicle counts}
\centering
\begin{tabular}{c||ccccc}
\hline
\textbf{Initial Vehicle Count} & 200 & 300 & 400 & 500 & 600 \\ \hline
\(\beta\)                      & 0.02 & 0.02 & 0.07 & 0.09 & 0.05 \\ \hline
\end{tabular}
\label{table:beta_values}
\end{table}

First, we examined whether our QUBO approach could reduce congestion more effectively than the conventional fixed-cycle control method, which cycles through predetermined signal durations. For this comparison, we adopted SUMO’s default configuration, where the cycle time for each intersection is set to 90 seconds, with equal signal durations assigned to each mode. For instance, if an intersection consists of two modes, each mode is allocated 45 seconds.

\begin{figure}[H]
\centering
\includegraphics[width=\columnwidth, keepaspectratio]{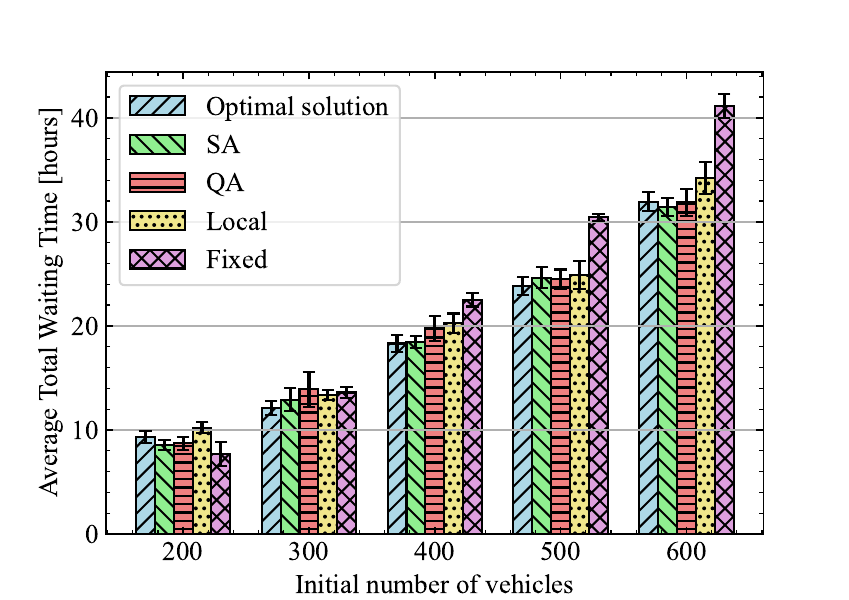}
\caption{
(Color online)
Comparison of total waiting time for different initial vehicle counts across multiple methods. The horizontal axis represents the initial number of vehicles, while the vertical axis shows the average total waiting time over 10 simulation instances, including standard error bars. "Optimal solution" refers to the solution computed by the Gurobi Optimizer. "Fixed" indicates results from a conventional fixed-cycle signal control. "Local" represents signal control without incorporating the second-order terms in Equation 6, while "QA" and "SA" show results from quantum annealing and simulated annealing, respectively. This comparison demonstrates the effectiveness of our QUBO approach, especially at higher traffic levels.
}

\label{fig:waiting_time}
\end{figure}

For each traffic flow level, we performed 10 simulation instances, measuring the \textit{total waiting time}, solution energy, and computation time. In each instance, simulations were run for 400 seconds with an optimization interval of $t_{interval} = 5$, resulting in 80 optimization calculations per instance. The Gurobi Optimizer was used to obtain the optimal solution, serving as a benchmark for evaluating the performance of QA and SA. First, we discuss the results for \textit{total waiting time} shown in Figure \ref{fig:waiting_time}. The horizontal axis represents vehicle flow, while the vertical axis indicates the mean and standard error of the \textit{total waiting time} over 10 instances. The case where signal timings are controlled by fixed-cycle scheduling is labeled as 'Fixed.' To assess the impact of $R_{im, jn}$ on congestion reduction, we also compare with a scenario where the optimal solution determines signal modes without including $H_2$ in Equation \ref{equ:final_cost}.

Table \ref{table:relative-error-waiting-time} shows the relative error between the optimal solution computed by Gurobi and the 'Fixed' cycle method. For an initial vehicle count of 200, the 'Fixed' method achieved a lower waiting time. However, as the number of vehicles increases, our proposed approach gradually reduces \textit{total waiting time}, reducing approximately 28.9\% at an initial vehicle count of 600. These results suggest that our method becomes more effective as vehicle density increases.

\begin{figure}[htbp]
\centering
\includegraphics[width=\columnwidth, keepaspectratio]{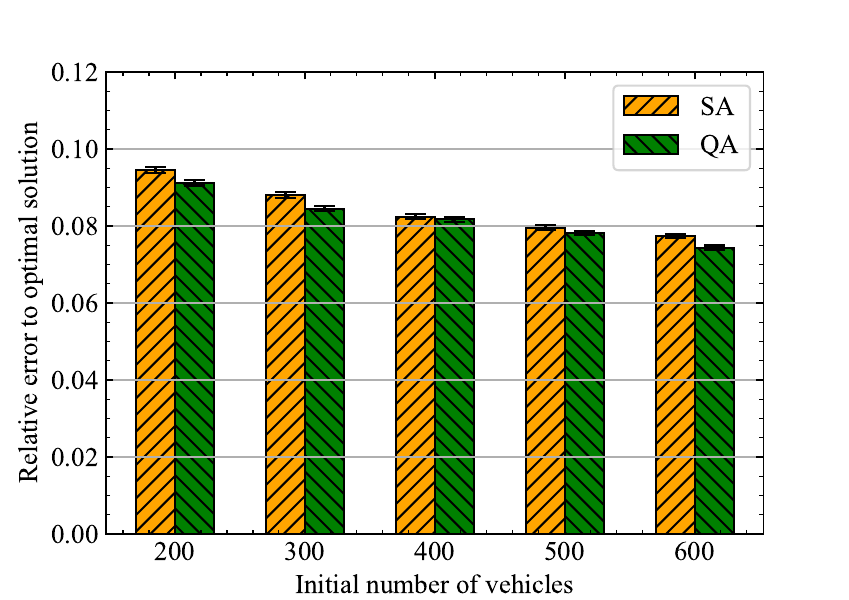}
\caption{
(Color online)
Relative error to the optimal solution for Simulated Annealing (SA) and Quantum Annealing (QA) across different initial vehicle counts. The horizontal axis shows the initial number of vehicles, and the vertical axis represents the relative error for the optimal solution from the Gurobi Optimizer. Both SA and QA exhibit relative errors of approximately 8-10\%, indicating that neither method consistently reaches the optimal solution. Each bar shows the mean relative error with standard error bars.
}
\label{fig:energy}
\end{figure}

Next, to evaluate the contribution of $H_2$, we compare the optimal solution with the 'Local' solution, revealing that the optimal solution consistently achieves a lower waiting time across all traffic flows. This result indicates that our proposed quadratic coefficient parameters help mitigate congestion more effectively, validating our QUBO formulation. In contrast, both QA and SA show a lower \textit{total waiting time} than the optimal solution at initial vehicle counts of 200 and 600. Figure \ref{fig:energy} illustrates the relative energy error between QA, SA, and the optimal solution. Across all traffic data, QA and SA failed to reach the optimal solution. Combined with the results shown in Figure \ref{fig:waiting_time}, this indicates that solutions with lower energy do not necessarily correspond to reduced \textit{total waiting time}. This result suggests that our cost function does not fully fit the true behavior of the \textit{total waiting time} function. This is likely because \textit{total waiting time} is a cumulative result derived from dynamically calculated solutions during each simulation, meaning that a fixed $ \beta $ cannot fit it perfectly in every scenario. This fit could improve if we could adjust $ \beta $ to the optimal value at each optimization step. However, such an approach would likely increase computational costs significantly and might lead to overfitting on specific traffic data. Thus, we propose adjusting $ \beta $ based on traffic flow, as this approach balances cost-effectiveness with performance. In practice, this approach allows for flexibility; for example, different $ \beta $ values could be applied during low traffic at night versus peak commuting hours.

\begin{table}[hbtp]
  \centering
  \caption{Relative Error in Total Waiting Time}
  \begin{tabular}{cccc}
    \toprule
    Initial number of vehicles & Optimal solution [hours] & Fixed [hours] & Relative error [\%] \\
    \midrule\midrule
    200 & 9.3 & 7.7 & -17.6 \\
    300 & 12.1 & 13.6 & 12.3 \\
    400 & 18.3 & 22.5 & 22.9 \\
    500 & 23.8 & 30.4 & 27.8 \\
    600 & 31.9 & 41.1 & 28.9 \\
    \bottomrule
  \end{tabular}
  \label{table:relative-error-waiting-time}
\end{table}

Returning to Figure \ref{fig:energy}, we see that QA consistently finds solutions with lower energy than SA. Although variables remain constant across initial vehicle counts, lower-energy solutions are more readily obtained as the vehicle count increases. This is likely because when vehicles are abundant on the map, the non-zero components in the first-order term increase, making the influence of these terms relatively stronger. In such cases, the dominance of the first-order terms reduces the impact of the more complex second-order terms, resulting in a cost function that is easier to solve. In contrast, when vehicle count is low, and many $ C_{i, m} $ coefficients are zero, the influence of the second-order terms becomes more significant, complicating the search for low-energy solutions. 

\begin{figure}[H]
\centering
\includegraphics[width=\columnwidth, keepaspectratio]{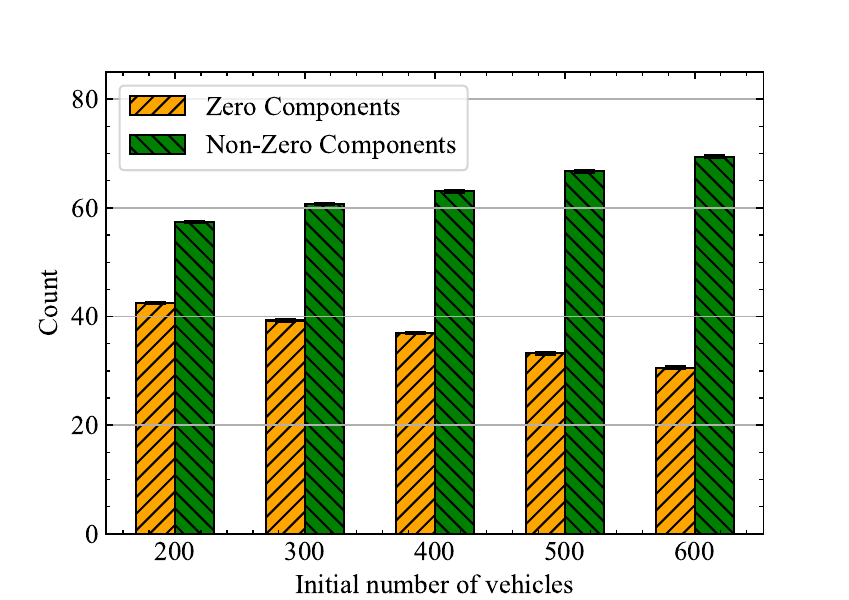}
\caption{
(Color online)
Counts of zero and non-zero components in the first-order terms across different initial vehicle counts. The horizontal axis represents the initial number of vehicles, while the vertical axis shows the count of zero and non-zero components. As vehicle count increases, the number of non-zero components rises, indicating that first-order terms become more influential under higher traffic conditions.
}
\label{fig:first_term}
\end{figure}

To substantiate this, we counted the number of zero and non-zero components in the first-order terms for each flow level. Figure \ref{fig:first_term} shows the mean and standard error of zero and non-zero component counts across 10 instances for each flow level. As expected, higher traffic flow correlates with more non-zero components in the first-order terms, suggesting that the QUBO formulation becomes easier to solve under these conditions.

\begin{figure}[htbp]
\centering
\includegraphics[width=\columnwidth, keepaspectratio]{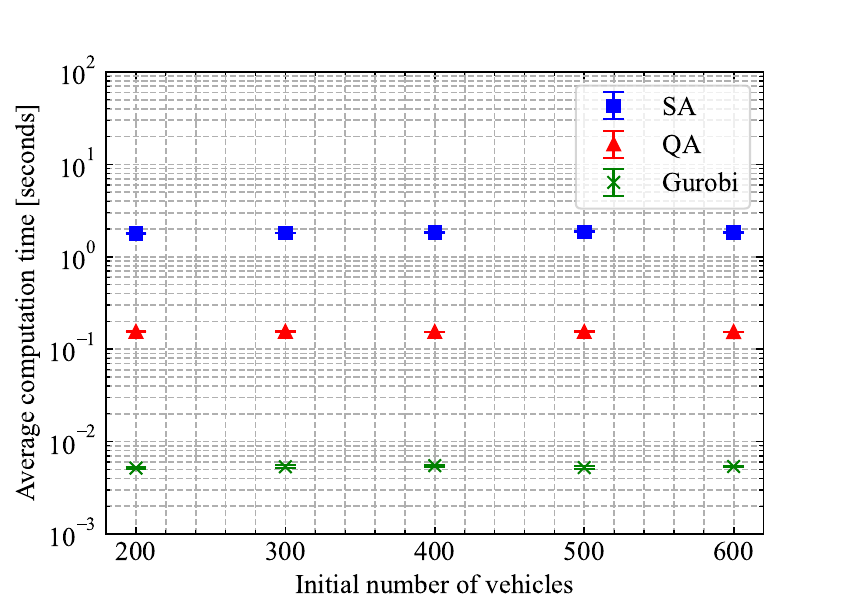}
\caption{
(Color online)
Average computation time for Simulated Annealing (SA), Quantum Annealing (QA), and the Gurobi Optimizer across initial vehicle counts. The horizontal axis represents the initial number of vehicles, and the vertical axis shows the average computation time on a logarithmic scale. Gurobi consistently achieves the shortest computation time, followed by QA and SA. Error bars indicate the standard error of the computation time across 10 simulation instances for each method.
}

\label{fig:calc_time}
\end{figure}

Finally, we compare computation times. Figure \ref{fig:calc_time} illustrates the mean and standard error of computation times for QA, SA, and Gurobi across 10 instances for each flow level. For Gurobi, we recorded the \texttt{Runtime} provided by the solver, while for SA, we logged the \texttt{sampling\_time} from OpenJij, and for QA, we obtained the \texttt{qpu\_access\_time} from the D-Wave Systems. The computations were performed using a 2022 MacBook Air with an Apple M2 chip and 16 GB of memory on macOS Sonoma 14.6.1. These results indicate that, at present, the Gurobi Optimizer computes the optimal solution most quickly. Although QA achieves lower energy faster than SA, it still lags behind Gurobi regarding computation time.

Based on these results, the current D-Wave machine neither achieves the optimal solution for our QUBO nor outperforms Gurobi in computation time. One potential reason for this outcome is the impact of chains generated during minor embedding, which increases the number of qubits used and adds complexity due to strong chain interactions, making it challenging to find the optimal solution. Additionally, internal parameters of the D-Wave machine may also affect performance. We used the default \texttt{annealing\_time} of 20 $\mu$s. Previous studies have demonstrated that longer \texttt{annealing\_time} settings and adjusted chain strengths can facilitate finding the optimal solution on D-Wave machines \cite{pelofske_comparing_2023}. 
Furthermore, changes to the \texttt{anneal\_schedule}, such as pausing the annealing process, can help obtain solutions with lower energy \cite{marshall_power_2019}. Thus, the performance of the D-Wave machine may be improved by adjusting these internal parameters.

The D-Wave machine also has a reverse annealing function. With this technique, the classical solution is used as the initial state, and a transverse magnetic field is gradually applied, starting from the classical state with the field turned off and then increasing and reducing it again. This approach allows the system to search for better solutions. A previous study demonstrated that reverse annealing facilitates reaching the ground state \cite{Haba2022}. Therefore, obtaining a low-energy solution using D-Wave machines requires careful parameter tuning and is more challenging than with a classical solver.

\section{Conclusions}
In this study, we formulated a QUBO under practical conditions and used SUMO to conduct simulations on actual map data, aiming to address the limitations of previous studies, which often conducted experiments under unrealistic conditions \cite{inoue_traffic_2021, hussain2020optimal}. We incorporated quadratic coefficients, $R_{im, jn}$, into the second-order terms to facilitate continuous passage through intersections. These coefficients are map-dependent and do not require recalculation at each optimization step, making them advantageous for traffic signal optimization, which demands fast dynamic calculations.

Our experimental results demonstrate that the proposed QUBO formulation effectively reduces \textit{total waiting time} compared to fixed-cycle control, particularly under high-traffic conditions. This outcome may be influenced by hardware limitations, such as the need for chains, and by insufficient tuning of internal parameters, such as \texttt{annealing\_time}. While future hardware developments may reduce the impact of chains, there will still be a need to automatically adjust internal parameters based on the specific QUBO.

The simulation results demonstrated that our QUBO formulation effectively reduced congestion under realistic conditions. However, the optimization interval in our simulations was set to 5 seconds, which may be slightly too long to achieve more responsive control. Reducing this interval may improve the effectiveness of optimization; however, excessively short intervals could lead to rapid mode changes, increasing the risk of traffic incidents. Additional modifications to the QUBO formulation are needed to facilitate optimization over shorter intervals. 
Potential approaches are discussed in the following section.

\section{Future Work}
In actual traffic signals, pedestrians have a minimum time to cross a pedestrian crossing safely. Frequent signal changes can lead to safety issues. 
If the optimization interval is shortened, Equation \ref{equ:H4} should be added to $H_1+H_2+H_3$.
\begin{equation}
\begin{aligned}
\label{equ:H4}
H_4(\vec{x}) &= 
\begin{cases} 
\sum_{i=1}^{N} \sum_{m \in M_i}x_{i m}\left(\tau_{im}-T_{i}\right)^2 & \text{if } \tau_{im} < T_i \\
0 & \text{if } \tau_{im} \geq T_i
\end{cases}
,
\end{aligned}
\end{equation}
where $\tau_{im}$ is the duration for which mode $m$ is selected continuously at the $i$-th intersection from when the mode is switched to the current time.
$T_{i}$ is the time required for a pedestrian to cross the pedestrian crossing at the $i$-th intersection. 

For example, consider the situation in which $T_{i}=10, \tau_{i0}=0, \tau_{i1}=7$ for the $i$-th intersection with two modes; that is, mode 1 is selected for 7 s in a sequence.
In this case, if mode 0 is selected, the cost increases by 100, and if mode 1 is selected, the cost increases by 9,
which causes mode 1 to be selected consecutively.

$T_{i}$ is determined by various factors, including the crosswalk length and traffic volume.
Because this information is difficult to obtain from the SUMO simulator, the experiment was conducted without considering $H_{4}$.
By extending the QUBO in this manner, the optimization timing can be set optionally.
Thus, our method can be used in the continuous-time domain, which allows for more realistic traffic signal control.

\begin{acknowledgment}
This work was supported by JSPS KAKENHI (Grant No. 23H01432).
This study was financially supported by the MEXT-Quantum Leap Flagship Program (Grant No. JPMXS0120352009), as well as the Public\verb|\|Private R\&D Investment Strategic Expansion PrograM (PRISM) and programs for Bridging the gap between R\&D and the IDeal society (society 5.0) and Generating Economic and social value (BRIDGE) from the Cabinet Office.
\end{acknowledgment}

\renewcommand{\refname}{
  References\\
  \small \textnormal{* reo.shikanai.s1@dc.tohoku.ac.jp}
}

\bibliographystyle{jpsj}
\bibliography{citation.bib}

\end{document}